\documentclass[12pt]{article}
\usepackage{graphicx}
\usepackage{color}

\newcommand {\eqref} [1] {(\ref {#1})}
\newcommand {\beq} {\begin{equation}} 
\newcommand {\eeq} {\end{equation}}
 \newcommand {\ber}{\begin{eqnarray*}}
 \newcommand {\eer} {\end{eqnarray*}}
\newcommand {\bea}{\begin{eqnarray}}
 \newcommand {\eea} {\end{eqnarray}}

\newcommand{\ket}[1]{|#1 \rangle}

\def\e{\epsilon}

\def\cZ{{\cal Z}}
 
\def\cN{{\cal N}}

\def\tr{{\rm Tr}}

\def\vphi{\varphi}

\def\nn{\nonumber}
\newcommand{\be}{\begin{equation}}
\newcommand{\ee}{\end{equation}}
\newcommand{\eq}[1]{(\ref{#1})}

\begin{document}

\hfill{NORDITA-2004-42}

\hfill{PAR-LPTHE 04-011}

\vspace{20pt}

\begin{center}

{\LARGE \bf $\cN = 2$ Super Yang--Mills and \\ the XXZ spin chain}

\vspace{30pt}
 
{\bf Paolo Di Vecchia$^{a}$,  
Alessandro Tanzini$^{b}$
}

\vspace{15pt}
{\small \em 
\begin{itemize}
\item[$^a$]
NORDITA,
Blegdamsvej 17, DK-2100 Copenhagen \O, Denmark
\item[$^b$] 
LPTHE, Universit\'e de Paris~VI-VII, 4 Place Jussieu 75252 Paris Cedex 05,
France. 
\end{itemize}
}

\vskip .1in {\small \sffamily divecchi@alf.nbi.dk,
tanzini@lpthe.jussieu.fr}
\vspace{50pt}

{\bf Abstract}
\end{center}
We analyse the renormalisation properties of composite operators
of scalar fields in the $\cN =2$ Super Yang--Mills theory.
We compute the matrix of anomalous dimensions in the planar limit
at one--loop order in the 't Hooft coupling, and show that it
corresponds to the Hamiltonian of an integrable XXZ spin chain
with an anisotropy parameter $\Delta>1$. 
We suggest that this parameter could be related to the presence
of non--trivial two--form fluxes in the dual supergravity
background.
We find that the running of the gauge coupling does not affect
the renormalization group equations for these composite operators
at one--loop order, and argue
that this is a general property of gauge theories which is not related to 
supersymmetry.  

\setcounter{page}0
\setcounter{footnote}0
\thispagestyle{empty}
\newpage

\section{Introduction}

The study of the AdS/CFT correspondence in the PP--wave limit
(for a review, see 
\cite{Maldacena:2003nj,Plefka:2003nb,Pankiewicz:2003pg,Sadri:2003pr,Russo:2004kr})
has triggered a great deal of work in the renormalisation properties of
composite operators in gauge theory. The original idea of Berenstein,
Maldacena and Nastase~\cite{Berenstein:2002jq} was to regard some
gauge--invariant operators of the $\cN=4$ Super Yang--Mills theory
having large $R$--charge as a discretised version of the physical type
IIB string on the PP--wave background.
The BMN operators are single
trace operators formed by a long chain of one of the elementary scalar
fields of $\cN=4$, with the insertion of a few other fields and
covariant derivatives (called {\it impurities}), each of them
corresponding to a different excitation of the string. The anomalous
dimensions of these operators is expected to coincide with the mass of
the corresponding string state~\cite{Berenstein:2002jq}. This matching
was first checked perturbatively at one-loop~\cite{Berenstein:2002jq}
and at two-loop~\cite{Gross:2002su} level, and then a field theory
argument was provided in~\cite{Santambrogio:2002sb} to extend the
correspondence to all orders of perturbation theory.

In~\cite{Gubser:2002tv} it was realised that the string theory 
states accessible by quantization
on the plane--wave background are indeed a subsector of a wider class 
of highly excited string states which can be described as 
semiclassical soliton solutions of the AdS$_5\times$S$^5$ string sigma model. 
The general feature of these states is that they carry large quantum numbers, 
corresponding to 
large angular momenta along the five sphere and/or the AdS space. 
The AdS/CFT dictionary enables one to 
identify the corresponding gauge theory operators.  
As in the BMN case, they are built as a long chain
of elementary fields, but in this case with an high number of impurities.
The computation of the anomalous dimensions of such operators is in general a
formidable task, due to the large number of different fields that they contain.

A very interesting observation was made in~\cite{Minahan:2002ve}, where
the matrix of the one--loop anomalous dimensions for the composite operators
of scalar fields of $\cN=4$ SYM theory in the planar limit  
was put in correspondence with the Hamiltonian of an integrable $SO(6)$
spin chain. This inspired further studies on the integrability of the planar
$\cN=4$ theory SYM at 
one--loop~\cite{Beisert:2003jj,Beisert:2003yb,Belitsky:2003sh,Dolan:2004ps}
and also at higher 
orders~\cite{Beisert:2003tq,Beisert:2003jb,Beisert:2003ys,Serban:2004jf,Ryzhov:2004nz,Beisert:2004hm}.
The relation with the integrable systems allows one to compute 
the anomalous dimensions of the ``long'' gauge theory operators
by using the algebraic Bethe ansatz.
In general, for states that have at least one large angular 
momentum $J$ along the five sphere, as for the plane--wave states,
one can define an effective expansion parameter $\lambda'=\lambda/J^2$.
In some cases, as for the BMN operators, both
the semiclassical expansion of string theory and the 
perturbative expansion in gauge theory can be defined in terms of
this effective parameter,
allowing for a quantitative comparison between the two. 
Several studies have been performed along these lines
and agreement has been found up to two loops\footnote{There is by now a huge
literature on this subject. An useful introduction together with a large
list of references can be found in the review
\cite{Tseytlin:2003ii}.}.
One salient feature of these developments is that they allow one 
to probe regions of the string spectrum far away from the states protected 
by supersymmetry. 
It is a remarkable fact that one finds a quantitative agreement 
also in these cases. 
It is thus conceivable that 
similar patterns of gauge/string duality can be unraveled also for theories
where some (or all) the supersymmetries are broken \cite{Kim:2003vn}
and the conformal invariance is lost 
\cite{Pons:2003ci,Alishahiha:2004vi,Bigazzi:2004yt}. 

Studies on the integrability in deformed $\cN=4$ SYM theories 
were performed in \cite{Roiban:2003dw}, which considered the Leigh--Strassler
deformation, in 
\cite{Wang:2003cu,Chen:2004mu}, where the $\cN=2,1$  
orbifold field theories were considered, and in \cite{DeWolfe:2004zt}
in the context of defect conformal field theories.
A general study on the integrability of $\cN=4$ SYM in presence of marginal
deformations has been recently presented in \cite{Berenstein:2004ys}.
 
In this paper, we focus our attention on the pure $\cN=2$ SYM theory, and we study
the renormalisation properties of composite
operators of scalar fields.
We find that the corresponding matrix of the anomalous dimensions 
reduces at one--loop and in the planar limit to
the Hamiltonian of an XXZ spin chain. 
We also study the renormalisation group flow for these
operators and show that the effects of the 
breaking of the conformal invariance show up only at the
two--loop order in the 't Hooft coupling.

\section{$\cN = 2$ theory and the XXZ spin chain}

We start by writing the Lagrangian of ${\cal{N}}=2$ Super Yang-Mills in
Weyl notations 
\begin{eqnarray}
L_{E} &=& \frac{2}{g^2} \tr\Big(\frac{1}{4}F_{\mu \nu}F_{\mu \nu} + (D_{\mu} 
\phi )^{\dagger} D_{\mu} \phi
+ \psi  \sigma^{\mu}  D^{\mu} \bar\psi 
+ \lambda \sigma^{\mu} D^{\mu}\bar\lambda
\nonumber\\
&& - i\sqrt{2}\left(
\psi [\bar\phi,\lambda] + \bar\psi  [\phi, \bar\lambda]
\right) 
+ \frac{1}{2}[\bar\phi,\phi]^2 \Big) \ ,
\label{lnee}
\end{eqnarray}
in terms of the euclidean $\sigma$-matrices 
$\sigma^{\mu}=({\bf 1},i\tau^i)$.
The field $\phi$ is the complex scalar of the $\cN=2$ Super
Yang-Mills, the two Weyl spinors $\lambda$ and $\psi$ are the
fermionic superpartners and the covariant derivative reads 
$D_{\mu} \phi = \partial_{\mu} \phi - i [A_{\mu}, \phi]$.

Let us start by studying the renormalization properties of operators
involving a product of the complex scalar field 
$\phi$~\footnote{We use the following
  conventions: the generators of the gauge group are normalized
  as $\tr (T^a T^b ) = \frac{1}{2} \delta^{ab}$
and the relations between the bare and renormalized quantities are
$g = Z_g g_r $ and $\phi  = Z_{\phi}^{1/2} \phi_r$.}
\begin{equation}
G_{J}(x_1,\ldots,x_J;z)=\cZ_{\phi}^{J/2} 
\langle\bar\phi_r (x_1)\ldots \bar\phi_r (x_J) 
\cZ_{{\cal O}} 
\tr(\phi^J)_r (z)\rangle=\cZ_{\phi}^{J/2}\cZ_{{\cal O}}
G_J^{(ren)}(x_1,\ldots,x_J;z)
\label{reno}
\end{equation}
$\cZ_{{\cal O}}$ is the renormalization factor for the composite
operator, and $\cZ_{\phi}$ is the usual wave--function renormalization
needed to make finite the two--point function 
$\langle\bar\phi_r (x)\phi_r (y)\rangle$. Notice that   the product of
the fields ${\bar{\phi}} \equiv \bar\phi^a (T^{a})_{ij}$  in the 
Green function (\ref{reno}) should
be understood as a product of the gauge group matrices. This means
that the Green function is a matrix, but sometimes we will not write
explicitly its indices in order to avoid a too heavy formalism.  
The factor $\cZ_{{\cal O}}$ is defined in order to reabsorb all the 
divergencies
arising in the computation of the bare correlator $G_J$ in \eq{reno}
with the Lagrangian \eq{lnee}.
It turns out that in ${\cal{N}}=2$ Super Yang-Mills all the self-energy 
diagrams cancel. This means that in the convention we choose for the
lagrangian \eq{lnee} the only renormalization for the fields is that associated
to the gauge coupling, {\it i.e.} $\cZ_{\phi}^{1/2}\equiv\cZ_g$.
On the other hand from the knowledge of the 
$\beta$-function
\begin{equation}
\beta(g) \equiv \mu\frac{\partial}{\partial\mu}g_r = 
-g \mu\frac{\partial}{\partial\mu} \log \cZ_g =
-\frac{g^3}{16\pi^2}2N
\label{beta2}
\end{equation}
one can derive the expression for $\cZ_g$
\begin{equation}
\cZ_g  = 1 - \frac{g^2 N}{16\pi^2} \frac{\mu^{-2\epsilon}}{\epsilon} \ .
\label{zg2}
\end{equation}
Indeed \eq{beta2} follows from \eq{zg2} after taking the $\epsilon\to 0$ limit.

We are now ready to study the renormalization properties of the
composite operator in \eq{reno}.
For the sake of clarity we start by considering the case $J=2$ 
\begin{equation}
G_{2} (x,y;z) \equiv 
\langle ( \bar\phi(x)\bar\phi(y) )_{ij} \tr(\phi^2)(z)\rangle \ .
\label{corr1}
\end{equation}
The tree level contribution in the planar limit 
is
\begin{equation}
G_{2} (x,y;z)|_{tree}
= \langle (\bar\phi(x) \bar\phi (y))_{ij} \tr(\phi^2)(z)\rangle_{\rm tree}
=g^4 \frac{N}{2} \Delta_{xz}\Delta_{yz} \delta_{ij}
\label{tree}
\end{equation}
where we have used the scalar field propagator
\begin{equation}
\langle {\bar{\phi}}_{ij} (x)   \phi_{hk} (y)    \rangle
 = \frac{g^2}{2}   \left(  \delta_{ik} \delta_{jh}\ -
 \frac{1}{N} \delta_{ij} \delta_{hk} \right)
\Delta_{xy}
\label{propa42}
\end{equation}
with
\begin{equation}
\Delta_{xy} \equiv 
\int \frac{d^{2\omega}p}{(2\pi)^{2\omega}}
\frac{e^{ip\cdot (x -y)}}{p^2} =  
\frac{\Gamma(\omega-1)}{4\pi^\omega(|x-y|^2)^{\omega-1}} \ .
\label{propa43}
\end{equation}
We use dimensional regularization with the dimension of the
space-time equal to $2 \omega \equiv 4 - 2 \epsilon  $.
At one-loop  the previous correlator has two contributions: one coming
from  the scalar potential and the second coming from the gluon exchange.
In principle there could be also the contribution of the self-energy
diagrams that, however, cancels as we have already remarked. 

\begin{figure}[ht]
\begin{center}
{\scalebox{1}{\includegraphics{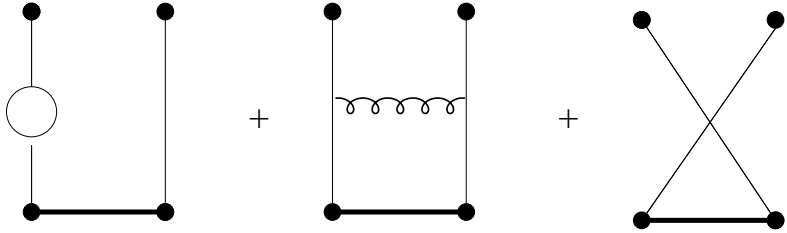}}}
\caption{Feynman diagrams contributing at one--loop. The thick horizontal
line joins the fields belonging to the composite operator. 
\label{f1}}
\end{center}
\end{figure}

\noindent
In the planar limit we get 
\begin{equation}
G_{2} (x,y;z)|_{1-loop;planar} = G_{2} (x,y;z)|_{tree} \times g^2 N
\left(K(z;x,y) + {\cal G}(z;x,y) \right)
\label{corr34}
\end{equation}
where
\begin{equation}
K(z;x,y) \equiv \Delta_{xz}^{-1} \Delta_{yz}^{-1}
\int d^{2 \omega} u \Delta_{xu}(\Delta_{zu})^2\Delta_{yu}  \sim 
\frac{1}{16 \pi^2} \left(\frac{1}{\e} + \dots
\right) 
\label{kzxy}
\end{equation}
and
\[
{\cal G}(z;x,y)=\Delta^{-1}_{zy}\Delta^{-1}_{zx}
\int d^{2\omega}y_1 \int d^{2\omega}y_2
(\Delta_{zy_1}\stackrel{\leftrightarrow}{\partial_\mu}^{y_1}\Delta_{y_1
x})
(\Delta_{zy_2}\stackrel{\leftrightarrow}{\partial_\mu}^{y_2}\Delta_{y_2
  y})
\Delta_{y_1y_2} \sim 
\]
\begin{equation}
\sim \frac{1}{16 \pi^2} \left(\frac{1}{\epsilon} +     \dots \right) \ .
\label{gzxy}
\end{equation}
The symbol $\stackrel{\leftrightarrow}{\partial_\mu}$ in \eq{gzxy}
stands for the left--right derivative 
$(A\stackrel{\leftrightarrow}{\partial_\mu}B)=A\partial_\mu B -\partial_\mu AB$.
The fastest way to get the planar contribution in the correlator
containing the string of fields ${\bar{\phi}}$ and a product of
various traces is first to perform all the possible contractions that
reduce everything to a single string of fields and then contract only
the fields that are next to each other in this single string. This
procedure maximizes the number of factors  $N$ and gives the planar
contribution. 
Both in (\ref{kzxy}) and (\ref{gzxy}) we have kept only the divergent
terms that are the ones that we need, while the dots represent the finite parts.
Eq.(\ref{corr34}) can be easily generalized, at the planar level, to
arbitrary $J$  
\bea
&&G_{J} (x_1\ldots x_i,x_{i+1}\ldots x_J;z)|_{1-loop;planar} = \nn \\
&&
=G_{J} (x_1\ldots x_i,x_{i+1}\ldots x_J;z)|_{tree;planar} \sum_{i=1}^J
\frac{g^2 N}{2} \left[K(z;x_i,x_{i+1}) + {\cal G}(z;x_i,x_{i+1})
\right] \sim \nn \\
&&\sim  G_{J} (x_1\ldots x_J;z)|_{tree} \times g^2 N
\frac{J}{16 \pi^2} \frac{\mu^{- 2 \epsilon}}{\epsilon}
\label{corrj}
\eea
where $(x_i,x_{i+1})$ are two nearest--neighbors fields in the
correlator and
\begin{equation}
G_{J}  ( x_1 , x_2, \dots x_J ;z )_{tree,planar}  = \frac{J N^{J-1}
  g^{2J}}{2^J } \delta^{ij} \Delta_{x_1 z} \Delta_{x_2 z} \dots \Delta_{x_J z}
\label{planatree}
\end{equation}
In the last term of \eq{corrj} we wrote the divergent terms, coming
half from the gluon exchange 
and half from the four scalar interaction. 
Collecting the tree--level and the one loop contributions together we have
\be
G_{J} (x_1\ldots x_J;z)_{one-loop;planar}=  G_{J} (x_1\ldots
x_J;z)|_{tree;planar} 
\left( 1 +  g^2 N \frac{J}{16 \pi^2} \frac{\mu^{- 2 \epsilon}}{\epsilon}
\right)
\label{tre+uno}
\ee
Finally, by converting the bare coupling factor $g^{2J}$ appearing in 
$G_J|_{tree}$ into the renormalised coupling $g_r^{2J}$, \eq{tre+uno} gets 
a factor $\cZ_g^{2J}$. 
If we now consider the renormalized correlator in \eq{reno}, by using
\eq{zg2}
we see that the divergence appearing in \eq{tre+uno} is exactly 
cancelled  by the
renormalization factors $\cZ_{\phi}^{-J/2}\cZ_g^{2J}=\cZ_g^J$, 
without the need of any  renormalization constant for
the composite operator, {\it i.e.} $\cZ_{{\cal O}} =1$.  This means
that these operators
are protected at one--loop. 
Indeed, in \cite{Blasi:2000qw,Lemes:2000db,Lemes:2000ni,Maggiore:2001zw}
it was shown that the generalised 
Slavnov--Taylor identities associated to the $\cN=2$ supersymmetry imply  
that these operators have vanishing anomalous dimensions
to all orders in perturbation theory. The above computation
provides an explicit check of this property at one--loop.

Let us now come to the more general case of composite operators of the two
real scalar fields of the $\cN=2$ theory
\be
{\cal O}=\tr\left(\varphi_{i_{1}}\ldots \varphi_{i_{l}}
\varphi_{i_{l+1}}\ldots\varphi_{i_L}\right) \ ,
\label{O-real}
\ee
where
\begin{equation}
\phi = \frac{1}{\sqrt{2}} \left(\varphi_1 + i \varphi_2 \right) \ .
\label{real7}
\end{equation} 
We derive the one-loop planar mixing matrix 
for anomalous dimensions following very closely the procedure used
in~\cite{Minahan:2002ve} for ${\cal{N}}=4$. We then study
the correlator
\be
\langle \varphi^{i_L}\ldots\varphi^{i_{l+1}}\varphi^{i_l}\ldots\varphi^{i_1}
{\cal O}\rangle= \cZ_\varphi^{L/2}\cZ_{\cal O} 
\langle \varphi^{i_L}_r\ldots\varphi^{i_{l+1}}_r\varphi^{i_l}_r
\ldots\varphi^{i_1}_r
{\cal O}_r\rangle
\label{corr-real}
\ee
where $\varphi=\cZ_\varphi^{1/2}\varphi_r$ and $\cZ_\varphi=\cZ_\phi$
by virtue of \eq{real7}. 
The operators (\ref{O-real}) mix among themselves at the quantum
level, and ${\cal Z}_{\cal O}$ is a matrix carrying the indices of the real fields.
One can wonder whether the set of operators (\ref{O-real}) is closed under renormalization. 
First of all, one can easily 
see that the one--loop diagrams in which two real fields of the operator (\ref{O-real}) combine 
to emit a gluon or a fermionic current are vanishing for symmetry reasons.
Then the only remaining possibility 
is a mixing with scalars operators containing derivatives. 
These operators do indeed appear in the counterterms needed for the
renormalisation of the operators (\ref{O-real}) \cite{Bianchi:2003eg}.
However, the converse is not true, since the operators (\ref{O-real})
do not appear as counterterms in the one--loop renormalisation
of operators containing derivatives. This implies that the 
mixing matrix is triangular and one can disregard the mixing
with derivative operators as far as the computation of the one--loop
anomalous dimensions is concerned.
We need then to study only the correlators (\ref{corr-real}).
By using the large $N$ approximation, we focus  on the 
nearest--neighbors interaction
\begin{equation}
\langle\ldots\vphi_{i_{l+1}}(x)\varphi_{i_l}(y)
\ldots\tr\Big(\ldots\vphi_{j_l}\varphi_{j_{l+1}}\ldots\Big)(z)\rangle
\label{nn-real}
\end{equation}
The one--loop correction associated to the gluon exchange is exactly the same
as that for the complex fields, and can be read from the last line of 
\eq{corrj}
by taking only half the contribution as explained just after \eq{corrj}. 
By introducing
also the (diagonal) index structure of the real scalar fields, we get
\begin{equation}
\cZ^{(gluon)\ldots j_l j_{l+1}\ldots}_{\ \ \ldots i_l i_{l+1}\ldots}= {\bf 1}  
+ \frac{g^2N}{16\pi^2}\frac{\mu^{-2\epsilon}}{2\epsilon}\
\delta_{i_l}^{j_l}\delta_{i_{l+1}}^{j_{l+1}} \ \ . 
\label{Z-gluon}
\end{equation}
In order to compute the contribution of the four-scalar interaction, it is
convenient to rewrite it in terms of real scalars
\begin{equation}
V_4=
- \frac{1}{2 g^2} \sum_{i,j =1}^{2} \tr \left( [ \varphi_i , \varphi_j ] 
[ \varphi_i , \varphi_j ] \right) 
\label{4scalar}
\end{equation}
The correction associated to four scalar interaction \eq{4scalar} is 
\begin{eqnarray}
&&\langle\vphi_{i_{l+1}}(x)\vphi_{i_l}(y)
\times \nonumber\\
&&\int d^{2\omega}u 
\frac{1}{2g^2}\tr\Big[\sum_{l,m=1}^2 \Big(2\vphi_l\vphi_m \vphi_l\vphi_m
-\vphi_l\vphi_m\vphi_m\vphi_l -\vphi_l\vphi_l\vphi_m\vphi_m\Big)\Big](u)
\tr\Big(\vphi_{j_l}\vphi_{j_{l+1}}\Big)(z)\rangle
\nonumber \\
&&
= g^4\frac{N}{2^2}\Delta_{xz}\Delta_{yz}\times
\frac{g^2}{2}\frac{N}{2^2}\ 4 K(z;x,y) 
\Big( 2 \delta_{i_l}^{j_{l+1}}\delta^{j_l}_{i_{l+1}}
- \delta_{i_l}^{j_{l}}\delta_{i_{l+1}}^{j_{l+1}}
- \delta_{i_l i_{l+1}}\delta^{j_l j_{l+1}} \Big)
\nonumber \\
&&\sim  g^4\frac{N}{2^2}\Delta_{xz}\Delta_{yz} \times 
\frac{g^2N}{16\pi^2}\frac{\mu^{-2\epsilon}}{2\epsilon}\
\Big( 2 \delta_{i_l}^{j_{l+1}}\delta^{j_l}_{i_{l+1}}
- \delta_{i_l}^{j_{l}}\delta_{i_{l+1}}^{j_{l+1}}
- \delta_{i_l i_{l+1}}\delta^{j_l j_{l+1}} \Big) \ ,
\label{four-real}
\end{eqnarray}
The factor $N/2^2$ appearing in the second line of \eq{four-real}
is associated to the matrix contractions 
and the factor $4$ comes from the four possible contractions with
the fields in the vertex.
The $\cZ$ factor associated to the above correction is
\begin{equation}
\cZ^{(four \ sc.)\ldots j_l j_{l+1}\ldots}_{\ \ 
\ldots i_l i_{l+1}\ldots}= {\bf 1}  
+ \frac{g^2N}{16\pi^2}\frac{\mu^{-2\epsilon}}{2\epsilon}\
\Big( 2 \delta_{i_l}^{j_{l+1}}\delta^{j_l}_{i_{l+1}}
- \delta_{i_l}^{j_{l}}\delta_{i_{l+1}}^{j_{l+1}}
- \delta_{i_l i_{l+1}}\delta^{j_l j_{l+1}} \Big) \ .
\label{Z-four}
\end{equation}
In conclusion the contributions coming from the gluon exchange and
from the four-scalar interaction are the same as in the ${\cal{N}}=4$
case~\cite{Minahan:2002ve} except that now the indices $i_l , i_{l+1} , j_l ,
j_{l+1}$ run only over two values and not six because ${\cal{N}}=2$ Super
Yang-Mills has only two real scalars. In addition in ${\cal{N}}=4$ we
have also the contribution of the self-energy diagrams. 
In the ${\cal{N}}=2$ case, as we already remarked, there are instead no self--energy corrections
and the renormalization of the fields is given at one--loop by
the coupling constant $\cZ_g$--factor. More precisely, in \eq{corr-real} it appears  
the factor $\cZ_\varphi^{L/2}=\cZ_g^L$. To pass from the 
bare coupling $g^{2L}$ appearing in the bare correlator in \eq{corr-real} to the renormalised one
we still have to multiply the r.h.s. of \eq{corr-real} by a $\cZ_g$ factor for each field.
This amount to the following renormalisation factor for the nearest--neighbors
\begin{equation}
 \cZ^{(g)\ldots j_l j_{l+1}\ldots}_{\ \ \ldots i_l i_{l+1}\ldots}= {\bf 1}  
- \frac{g^2N}{8\pi^2}\frac{\mu^{-2\epsilon}}{2\epsilon}\
\delta_{i_l}^{j_l}\delta_{i_{l+1}}^{j_{l+1}} \ \ . 
\label{Z-g}
\end{equation}
Adding the three contributions in (\ref{Z-gluon}),
(\ref{Z-four}) and (\ref{Z-g}) we get
\begin{equation}
{\cal{Z}}^{\ldots j_l j_{l+1}\ldots}_{\ \ \ldots i_l i_{l+1}\ldots}= {\bf 1} -
\frac{g^2 N}{16 \pi^2} \frac{\mu^{-2\epsilon}}{2\epsilon} \left( 
\delta_{i_l i_{l+1}}\delta^{j_l j_{l+1}}  + 2 
\delta_{i_l}^{j_l}\delta_{i_{l+1}}^{j_{l+1}} - 
 2 \delta_{i_l}^{j_{l+1}}\delta^{j_l}_{i_{l+1}}\right) \ .
\label{anodim}
\end{equation}
The resulting matrix of anomalous dimensions for these operators 
turns out to be the same as in
$\cN=4$ theory \cite{Minahan:2002ve}, except that now the indices run
only over the values $1$ and $2$~\footnote{Another difference is that 
 the 't Hooft coupling that appears in 
 \eq{anodim} is the renormalised running coupling
 $\lambda_r=g_r^2N$. However, the substitution $\lambda\to\lambda_r$
 induces only higher order corrections. With this remark in mind,
 we will write our results in terms of the bare coupling
 $\lambda$ to simplify the notation.}. 
For the particular case of the composite operators of complex fields
in \eq{reno}, the matrix is vanishing, in agreement with the result
discussed after \eq{corrj}. In fact these operators, when represented
in terms of the real fields $\varphi_i$, are symmetric
and traceless in the real indices $i=1,2$, and this ensures the vanishing
of their one--loop anomalous dimensions computed from \eq{anodim}. In this sense, these operators 
are the analogous in the $\cN=2$ theory of the BPS (or chiral primary) operators
of the $\cN=4$ SYM \cite{Blasi:2000qw,Lemes:2000db,Lemes:2000ni,Maggiore:2001zw}.

As another example, one can consider the Konishi operator ${\cal K}=\tr\bar\phi\phi$.
In the direct computation of the planar, one--loop renormalisation of this operator
one finds that the contribution of the gluon exchange has the opposite sign
with respect to that of the four scalar interaction. Thus they cancel out and
the only renormalisation of the Konishi operator is that associated to the gauge
coupling: ${\cal Z}_{\cal K}=\cZ_g^2$. From this it follows that $\gamma_{\cal K}=\lambda/4\pi^2$.
On the other side, when one acts
on ${\cal K}$ with the matrix (\ref{anodim}), the contributions of the identity and of the permutation
operators compensate each other and only the trace contribution is left. By summing
on the two sites and using (\ref{zg2}) one gets again ${\cal Z}_{\cal K}=\cZ_g^2$, in agreement with 
the direct computation. 

Let us now come to the discussion of the relation with the spin chain. 
Quite naturally the two scalar
fields of the ${\cal N}=2$ SYM can be interpreted as different
orientations of a spin and then the whole gauge invariant operator
formed just by scalars can be seen as a spin chain. The cyclicity of
the trace makes the chain closed 
and implies that the physical states of the chain
corresponding to the gauge theory operators have zero total momentum.

\begin{figure}[ht]
\begin{center}
{\scalebox{1}{\includegraphics{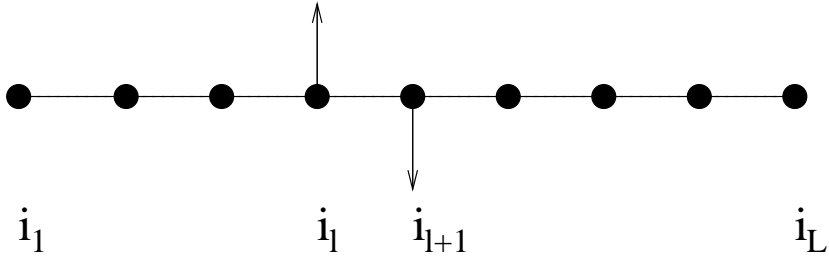}}}
\begin{picture}(0,0)%
\end{picture}%
\caption{Spin chain
\label{f2}}
\end{center}
\end{figure}

From this point of view, we can interpret the matrix of the one-loop
anomalous dimensions 
\be
\gamma_{\cal O} \equiv \cZ_{\cal O}^{-1} \mu \frac{\partial}{\partial\mu}\cZ_{\cal O}
\label{gamma-O}
\ee
which we get from \eq{anodim} 
\begin{equation}
\gamma_{\cal O} = \frac{g^2 N}{16 \pi^2} \sum_{l=1}^{L} \left(K_{l, l+1} +2 - 2
P_{l, l+1}\right)= \frac{\lambda}{16\pi^2}\sum_{l=0}^L H_{l,l+1}
\label{gaano}
\end{equation}
as a Hamiltonian acting on the spin chain.
Since the indices of the real scalar fields of the $\cN=2$ theory run just from one to two,
we can directly rewrite the matrix of anomalous dimensions \eq{gaano}
in the basis of sigma matrices 
$\hat\sigma_\mu\equiv ({\bf 1},\hat\sigma^x, \hat\sigma^y, \hat\sigma^z)$
which is better suited for the study  
of spin chain systems. In this basis $\varphi_1 \rightarrow (1\,\, 0)$
and $ \varphi_2 \rightarrow (0 \,\, 1)$. The permutation operator is
\begin{equation}
P^{ j_l j_{l+1} }_{i_l i_{l+1}} \equiv 
\delta_{i_{l}}^{j_{l+1}}\delta_{i_{l+1}}^{j_{l}}  =
\frac{1}{2} \sum_{\mu=0}^{3} \beta_{\mu} ( \hat\sigma^{\mu} )^{j_l}_{~~i_l}
( \hat\sigma^{\mu} )^{j_{l+1}}_{~~i_{l+1}}  \ ,
\label{pmat}
\end{equation}
with coefficients $ \beta_{\mu} = (1,1,1,1)$, and the ``trace''
contribution, where the two consecutive real fields are equal, is
\begin{equation}
K^{j_l j_{l+1}}_{  i_l i_{l+1}} \equiv 
\delta_{i_l i_{l+1}} \delta^{j_{l} j_{l+1}} =
\frac{1}{2} \sum_{\mu=0}^{3} \alpha_{\mu} ( {\hat{\sigma}}^{\mu} )^{j_l}_{~
i_l}  ( {\hat{\sigma}}^{\mu} )^{j_{l+1}}_{~~i_{l+1}}  
\label{kmat}
\end{equation}
where $ \alpha_{\mu} = (1, 1, -1, 1)$.
Moreover, before to write down the spin chain Hamiltonian we
observe that the operators containing only products
of the complex scalar field $\phi$ have vanishing anomalous dimensions.
This suggest to take them as the lowest energy eigenstates of the spin chain
and to identify the two orientations of a spin with the following
$2$-vectors
\beq
\bar\phi \to \ket{+} \equiv
\left(
\begin{array}{c}
1\\ 0
\end{array} 
\right) ~~,~~~~
\phi \to \ket{-} \equiv
\left(
\begin{array}{c}
0\\ 1
\end{array} 
\right)~.
\label{spin}
\eeq
From \eq{real7} one immediatly realises that
the change of basis required to satisfy \eq{spin} has to
exchange the two Pauli matrices $\hat\sigma^y\to -\sigma^z$ and $\hat\sigma^z\to \sigma^y$
leaving unchanged the third $\hat\sigma^x\to\sigma^x$.
The action of the Pauli matrices in the basis \eq{spin} is $\sigma^x \ket{+} =
\ket{-}$, $\sigma^x \ket{-} = \ket{+}$, $\sigma^y \ket{+} = i
\ket{-}$, $\sigma^y \ket{-} = -i \ket{+}$, $\sigma^z \ket{+} =
\ket{+}$ and $\sigma^z \ket{-} = - \ket{-}$, and
the matrix of anomalous dimensions \eq{gaano} reads
\bea
\gamma_{\cal O} &=& - \frac{g^2 N}{32 \pi^2} \sum_{l=1}^{L} \left[ 
(\sigma^x )_{l} ( \sigma^x )_{l+1} + 
(\sigma^y )_{l} ( \sigma^y )_{l+1} +
3 \left( (\sigma^z)_{l} ( \sigma^z )_{l+1} - {\bf 1}_l {\bf 1}_{l+1} \right)
\right] \nonumber\\
&\equiv& \frac{\lambda}{16\pi^2} H_{\rm XXZ} \ ,
\label{gaano34}
\eea  
where
\be
H_{\rm XXZ}= -\frac{1}{2} \sum_{l=1}^{L} \left[ 
(\sigma^x )_{l} ( \sigma^x )_{l+1} + 
(\sigma^y )_{l} ( \sigma^y )_{l+1} +
3 \left( (\sigma^z)_{l} ( \sigma^z )_{l+1} - {\bf 1}_l {\bf 1}_{l+1} \right)
\right]
\label{xxz}
\ee
is the Hamiltonian of an XXZ spin system!!\cite{Baxter}.  
An
interesting feature of this system is that it displays an anisotropy
parameter $\Delta$.
The behaviour of the spin chain depends critically on the value of this parameter; in
particular for $\Delta>1$ the spectrum has a mass gap \cite{Baxter}. 
The integrable system that we find in the $\cN=2$ case belongs to this class, 
since from (\ref{xxz}) we read $\Delta=3$. As anticipated, the ground state
of the spin chain corresponds to the protected operator ${\cal O}_{vac}\equiv\tr(\phi^L)$.
The excited states are associated to spin flips along the chain,
which in the field theory language correspond to the insertion of ``impurities''
$\bar\phi$ in the operator ${\cal O}_{vac}$. 
Due to the constraint of zero total momentum imposed by the cyclicity
of the trace, we have to consider at least two impurities, {\it i.e.}
we study the operator 
\be
{\cal O}_{n}=\sum_{l=0}^L \omega_l
\tr\left(\bar\phi\phi^l\bar\phi\phi^{L-l}\right) \ .
\label{flip}
\ee
The chain being closed, the coefficients 
$\omega_l$ appearing in (\ref{flip})
are some periodic functions of the position $l$.
The anomalous dimensions of the operator (\ref{flip}) 
were computed in \cite{Berenstein:2002jq} for the $\cN=4$ SYM theory in the $L\to\infty$
limit.
In this limit, one can take $\omega_l={\rm e}^{2\pi{\rm i} n \frac{l}{L}}$.
We remark that the relevant diagrams for the computation of \cite{Berenstein:2002jq}
come from the D--term
interaction in the last line of (\ref{lnee}), and are exactly
the same for the ${\cal N}=2$ theory.
All the other diagrams
were effectively taken in account in \cite{Berenstein:2002jq}
by imposing that their contribution cancel when putting $n=0$ and
$\bar\phi\to\phi$, since the corresponding operator is protected.
This argument is still valid in the $\cN=2$ theory, where 
we have shown that the operator ${\cal O}_{vac}$ does not
get quantum corrections.
We can thus compute the value of the anomalous dimension
of the operator (\ref{flip}) from the XXZ Hamiltonian by using (\ref{gaano34})  
and directly compare it with the result of \cite{Berenstein:2002jq}.
By using the identification (\ref{spin}) and the action of the Pauli matrices on the sites recalled 
after \eq{spin}, one can easily get
\be
\gamma_{\cal O}{\cal O}_n = -\frac{\lambda}{32\pi^2}
\sum_{l=0}^L 4 (\omega_{l+1} + \omega_{l-1} - 2\omega_l - 4\omega_l + \ldots)
\tr\left(\bar\phi\phi^l\bar\phi\phi^{L-l}\right) \ ,
\label{g-flip}
\ee
where the factor $4$ is the number of links between $\phi$ and $\bar\phi$ in the operator ${\cal O}_{n}$, and
the dots represent subleading terms in the $L\to\infty$ limit.
In (\ref{g-flip}) we splitted $\Delta=1+2$ to evidentiate the contribution of the 
anisotropy. In fact, by using the explicit expression of $\omega_l$ we get
\bea
\gamma_{\cal O}{\cal O}_n &=& -\frac{\lambda}{8\pi^2} 
\left[({\rm e}^{2\pi{\rm i}\frac{n}{L}} - {\rm e}^{-2\pi{\rm i}\frac{n}{L}} - 2)
- 4\right]{\cal O}_n \nonumber\\
&\sim& \frac{\lambda}{8\pi^2} \left(4+ \frac{4\pi^2n^2}{L^2}\right){\cal O}_n \ ,
\label{bmn}
\eea
in agreement with the result of \cite{Berenstein:2002jq}~\footnote{In order to compare 
with Eq.(A.18) of \cite{Berenstein:2002jq} one has to rewrite \eq{bmn}
in terms of the string coupling $g_s$ by using $g^2=4\pi g_s$ and to divide by two 
since BMN considered the effect of a single impurity $\bar\phi$.}. 
We thus see that the presence of 
a non--trivial anisotropy parameter $\Delta>1$ implies the presence of a {\it mass gap}
of the order of the 't Hooft coupling $\lambda$ in the spectrum.
This behaviour of the anomalous dimensions led BMN to conclude
that the string states corresponding to these operators become very massive
in the $\lambda\to \infty$ limit and decouple from the spectrum of the free
string on the PP--wave background \cite{Berenstein:2002jq}.
The situation may be different here in the context of the $\cN=2$ theory.
In fact the known ``dual'' supergravity solutions 
\cite{Bertolini:2000dk,Polchinski:2000mx,Gauntlett:2001ps,Bigazzi:2001aj,DiVecchia:2002ks}
describe actually some aspects of
the ultraviolet behaviour of the $\cN=2$ field theory, where the 
't Hooft coupling is small, and in fact reproduce the
correct perturbative running of the gauge 
coupling~\cite{Bertolini:2000dk,Polchinski:2000mx,DiVecchia:2002ks}. 
For this reason, it is possible that by studying spinning strings
on the background given by those solutions one
would be able to describe the perturbative anomalous scaling
dimensions of the composite operators that we studied.
In this sense it is suggestive to think that the presence of the anisotropy
parameter $\Delta$ in the spin chain could be related to the non trivial flux
of the NS two--form which breaks the isometry of the supergravity solution down
to $SO(2)$ and makes the gauge coupling constant to run. 

\section{Renormalisation group flow and the breaking of conformal invariance}

The presence in the $\cN=2$ theory of a non--trivial beta function
allows for the study of the effects of the breaking of conformal invariance
on the relation with the integrable model. 
In order to investigate on this issue, it is particularly interesting
to study the renormalisation group equations for
the composite operators \eq{O-real}. 
As before, we start by studying the particular case of the protected operators of complex fields
appearing in \eq{reno}.
To this end, let us define the 1PI Green function\footnote{The one--loop connected Green functions $G_J$ that we are studying
receive contributions only from 1PI graphs, see Fig.1. Thus we can get
the corresponding 1PI functions $\Gamma_J$ by simply amputing the external legs.}  
\be
\Gamma_J(x_1,\ldots,x_J;z)^{(ren)}
\equiv g_r^{-2J} \Delta^{-1}_{x_1z}\ldots\Delta^{-1}_{x_Jz}
G_J(x_1,\ldots,x_J;z)^{(ren)} \ .
\label{1pi}
\ee
The renormalisation group equation for \eq{1pi}
reads in general
\be
\Big(\mu\frac{\partial}{\partial\mu} + \beta_g \frac{\partial}{\partial g}
-J \gamma_\phi + \gamma_{{\cal O}}\Big) \Gamma_J(x_1,\ldots,x_J;z)^{(ren)}=0
\label{rge}
\ee   
where $\gamma_\phi$ is the anomalous dimension of the $J$ fields $\bar\phi$ appearing
in $\Gamma_J^{(ren)}$, and $\gamma_{{\cal O}}$ is the anomalous dimension of the operator
inserted in the Green function. 
For $\cN=2$ at one--loop the anomalous dimension of the fields is
\be
\gamma_\phi=\mu\frac{\partial}{\partial\mu}\log\cZ_g = g^2N \frac{1}{8\pi^2}
\label{gamma-fields}
\ee
while in this particular case the anomalous dimension of the operator is zero since $\cZ_{\cal O}=1$. 
Then
\eq{rge} reads in the $\cN=2$ case at one-loop
\be
\Big(\mu\frac{\partial}{\partial\mu} + \beta_g \frac{\partial}{\partial g}
-J \gamma_\phi \Big) \Gamma_J(x_1,\ldots,x_J;z)^{(ren)}=0
\label{rge-2}
\ee 
It is easy to verify this equation from the explicit computations that we already done.
In fact from \eq{corrj} we have
\bea
\Gamma_J(x_1,\ldots,x_J;z)^{(ren)}|_{1loop;planar}&=&\sum_{i=1}^{J}
\Big( \frac{g^2N}{2}[K(z;x_i,x_{i+1})+{\cal G}(z;x_i,x_{i+1})] -  
 \frac{g^2N}{16\pi^2}\frac{\mu^{-2\epsilon}}{\epsilon}\Big) \nonumber\\
&\sim& \frac{g^2N}{32\pi^2} \sum_{i=1}^J[\log(z-x_i)^2\mu^2 + \log(z-x_{i+1})^2\mu^2 ]
\label{gj-ren}
\eea
where the last term in the first line is the counterterm needed to
cancel the divergences
of the one--loop integrals. From \eq{gj-ren} we read the explicit $\mu$ dependence
\be
\mu\frac{\partial}{\partial\mu} 
\Gamma_J(y_1,\ldots,y_J;x)^{(ren)} = g^2N \frac{J}{8\pi^2}
\label{mu-gj-ren}
\ee
By using \eq{gamma-fields} one can see that this dependence is exactly canceled by 
the term $(-J\gamma_\phi)$ due to the anomalous dimensions of the $\bar\phi$ fields,
while the term associated to the beta function gives contribution only at higher orders,
starting from $\lambda^2$. 

The renormalization group equation \eq{rge-2} can be easily generalized to
the case of composite operators of real 
fields (\ref{O-real}). In this case, we can write 
\be
\Big(\mu\frac{\partial}{\partial\mu} + \beta_g \frac{\partial}{\partial g}
-J \gamma_\varphi \Big) \langle 
O^{(ren)}\cdot\Gamma\rangle= - \frac{\lambda}{16\pi^2} \sum_{l=0}^L H_{l,l+1} 
\langle O^{(ren)}\cdot\Gamma\rangle
\label{rge-O}
\ee 
where 
we used (\ref{gaano}) to write the anomalous dimensions of the operators (\ref{O-real}).
In (\ref{rge-O}) 
the symbol $\langle O^{(ren)}\cdot\Gamma\rangle$ stands for the insertion of the composite operator 
\eq{O-real}
in a generic 1PI Green function, and $\gamma_\varphi=\gamma_\phi$
is the anomalous dimension of the real scalar fields. The 
relevant point is that also in this case
the contribution of the term associated to the $\beta_g$--function 
is of order $g^2 N^4=\lambda^2$ and thus does not contribute to the renormalization
group equations \eq{rge-O} at one--loop order.
Thus the effects of the breaking of the conformal invariance for these operators
starts only at two--loops. 
This is simply due to the fact that the tree--level contribution to the 1PI Green
functions is $g$--independent, and that the $\beta_g$--function contribution
is proportional to $g^3$. These features are obviously valid in {\it any} gauge theory,
and thus one can argue that this behaviour is mantained also in non--supersymmetric
theories. In fact similar integrable systems have been known for some times
also in QCD \cite{Lipatov:1994yb,Faddeev:1995zg,Braun:1998id,Braun:1999te,
Belitsky:1999bf,Belitsky:2003ys, Ferretti:2004ba}.
A review on the use of the conformal symmetry in QCD phenomenology can be found
in \cite{Braun:2003rp}.

\section{Discussion}

In this paper we have shown that the one--loop renormalisation 
properties of composite operators in $\cN=2$ SYM theory are
related to the dynamics of an XXZ closed spin chain \cite{Baxter}.
Differently from the integrable systems usually discussed in the
context of $\cN = 4$ theory, like the XXX Heisenberg spin chain,
this dynamical system has an anisotropy parameter $\Delta$ which is responsible
for some interesting new properties.
The relation with the XXZ spin chain indicates that the integrable structure arising in
$\cN=2$ SYM is related to quantum groups differents from the Yangians appearing
in the $\cN=4$ theory \cite{Dolan:2004ps}. Moreover,
we have found that for the $\cN=2$ theory the XXZ Hamiltonian has an anisotropy
parameter 
$\Delta>1$. In this regime, the spectrum of the XXZ spin chain
displays a mass gap. We computed the energy of the first excited state (with
zero total momentum) in the limit of a long chain, and shown that it agrees
with the field theory results.     
Since the Bethe ansatz for the XXZ chain is known, it would be interesting to
apply it to compute the anomalous dimensions of gauge theory operators
of finite size and with a higher number of impurities.

We have also shown that the ground state of the XXZ spin chain corresponds to
symmetric traceless operators which are protected at one--loop.
These operators are the analogues in the $\cN=2$ theory of the BPS (or chiral
primary) operators of $\cN=4$, and were studied 
in~\cite{Blasi:2000qw,Lemes:2000db,Lemes:2000ni,Maggiore:2001zw} 
by using generalised Slavnov--Taylor identities related to the 
$\cN = 2$ supersymmetry. 
These identities imply the 
vanishing of the anomalous dimensions of these operators
to all orders of perturbation theory.
One can thus wonder whether the integrability properties found 
at one--loop can be extended to higher orders as well.

Concerning the breaking of the conformal invariance, we have
seen that the presence of a non--trivial beta function
does not modify the renormalisation group flow
of the composite operators at the leading order.
This seems to be a rather general feature of gauge
theories, not related to the presence of supersymmetry.
Similar relations with integrable models have in fact been
found also in large $N$ QCD~\cite{Lipatov:1994yb,Faddeev:1995zg,
Braun:1998id,Braun:1999te,Belitsky:1999bf,Belitsky:2003ys,Ferretti:2004ba}.
An unified framework for $\cN$--extended
supersymmetric theories with $0\le\cN\le 4$ 
has been recently proposed
in~\cite{Belitsky:2004yg} in the light--cone quantization. 
These features make particularly interesting to investigate
whether some relation can be found  
between the integrability of some subsectors
of gauge theories in the large $N$ limit 
and the existence of a dual string theory description for them.
One interesting direction would be to investigate the continuum limit of the
XXZ spin chain in the same spirit of the analysis performed in 
\cite{Kruczenski:2003gt,Kruczenski:2004kw,Dimov:2004qv,Kazakov:2004qf} 
for the XXX chain in $\cN=4$ theory and in \cite{Belitsky:2003ys,Ferretti:2004ba} for QCD.

\section*{Acknowledgements}
We thank R. Russo for collaboration at various stages of the project
and for many valuable comments.
We would like also to thank O. Babelon, G. D'Appollonio and  S.P. Sorella
for very useful discussions. A.T. is supported by Marie Curie fellowship of the
European Union under RTN contract HPRN-CT-2000-00131. 
This work is partially supported by the European
Commission RTN programme HPRN-CT-2000-00131.

\end{document}